\documentclass[10pt]{bmc_article}    

\usepackage{cite} 
\usepackage{url}  
\usepackage{ifthen}  
\usepackage{multicol}   
\usepackage{amssymb}
\usepackage{amsfonts}
\usepackage{amsmath}
\usepackage{graphicx}
\usepackage{subfigure}
\newboolean{publ}

\newenvironment{bmcformat}{\begin{raggedright}\baselineskip20pt\sloppy\setboolean{publ}{false}}{\end{raggedright}\baselineskip20pt\sloppy}

\begin{document}
\begin{bmcformat}
\title{Efficient mitigation strategies for epidemics in rural regions}
      \author{Caterina Scoglio$^{a}$ \footnote{Corresponding author:\\ Caterina Scoglio\\
Electrical and Computer Engineering, Kansas State University\\
2069 Rathbone Hall, Manhattan, Kansas 66506-5204\\
Tel: +1 785 532 4646, Fax: +1 785 532 1188\\
E-mail: caterina@ksu.edu}
      \and
         Walter Schumm$^{b}$
      \and
         Phillip Schumm$^{a}$
      \and
         Todd Easton$^{c}$  
      \and
         Sohini Roy Chowdhury$^{a}$    
      \and
         Ali Sydney$^{a}$
      \and 
         Mina Youssef$^{a}$}


\address{%
    \iid(a)Department of Electrical and Computer Engineering, Kansas State University, Manhattan, Kansas, USA-66506\\
    \iid(b)Department of Family Studies and Human Sciences , Kansas State University, Manhattan, Kansas, USA-66506 \\
    \iid(c)Department of Industrial and Manufacturing Systems Engineering, Kansas State University, Manhattan, Kansas, USA-66506
}

\maketitle


\begin{abstract} 
Containing an epidemic at its origin is the most desirable mitigation. Epidemics have often originated in rural areas, with rural communities among the first affected. Disease dynamics in rural regions have received limited attention, and results of general studies cannot be directly applied since population densities and human mobility factors are very different in rural regions from those in cities. We create a network model of a rural community in Kansas, USA, by collecting data on the contact patterns and computing rates of contact among a sampled population. We model the impact of different mitigation strategies detecting closely connected groups of people and frequently visited locations. Within those groups and locations, we compare the effectiveness of random and targeted vaccinations using a Susceptible-Exposed-Infected-Recovered compartmental model on the contact network. Our simulations show that the targeted vaccinations of only 10\% of the sampled population reduced the size of the epidemic by 34.5\%. Additionally, if 10\% of the population visiting one of the most popular locations is randomly vaccinated, the epidemic size is reduced by 19\% . Our results suggest a new implementation of a highly effective strategy for targeted vaccinations through the use of popular locations in rural communities.
\end{abstract}
\textbf{Keywords: }{complex networks, epidemics, survey, mitigation}

\section{Introduction}

Influenza A (H1N1), commonly known as swine flu, continues to be the dominant influenza virus in circulation across the globe with many countries and overseas territories reporting laboratory confirmed cases, including at thousands of deaths \cite{CDC2HF}. Factually, the origin of pandemic virus strains, such as the current H1N1, often trace back to rural regions. For example, the H1N1 2009 virus is suspected to have been originated in La Gloria, a small town near Veracruz, Mexico. Also, the previous strain of H1N1, commonly known as the Spanish Flu of 1918 that wrought devastation around the world, originated within the rural State of Kansas near Fort Riley. Other instances of epidemics originating in rural regions include the swine flu that originated in September 1988 at a hog barn in Walworth County, Wisconsin, the H5N3 virus that was identified at La Garnache farm in France in late January 2009, and the Asian flu that was a category 2 avian influenza in Ghizhou, China in 1956.

For analysis and containment purposes, large cities are generally considered to be infection hubs owing to the large population densities and mobility indices. Consequently, most spatio-temporal research on infectious human diseases focuses on large cities, such as Portland \cite{EGKMSTW}, Chicago \cite{V}, and Dresden \cite{PK}, which respectively represent excellent examples of an agent-based model, a multi-scale meta-population model, and a social-structure model, defining different levels of detail and complexity. Another approach to characterize the heterogeneous epidemic terrain of a human population is based on the construction of the network representing contacts among people. Studies of this type include \cite{PSV}, \cite{BPV}, and \cite{N2}.

Various immunization strategies have been formulated for urban populations. Some of these strategies assumed that the human population distribution can be estimated as scale free. One such strategy was a targeted immunization wherein nodes having the highest connectivity were deemed to be the most critical for spreading the
infection, and hence those highly connected people were chosen for vaccination \cite{PV}. However, global immunization strategies require the knowledge of the entire network of individuals and are complex from both computation and implementation stand points. Conversely, localized mitigation strategies, such as acquaintance immunization \cite{RSD}, randomly choose a subset of the entire population, and randomly select a set of their acquaintances to vaccinate. This strategies require a lower fraction of the population to be vaccinated than a random global immunization to dampen the impact of an epidemic. Other localized immunization methods include targeting acquaintances of randomly selected people by their estimated contact characteristics \cite{GLABH}.

Disease dynamics in rural regions \cite{Ketal} \cite{ZSJFRWBMN} have received limited study. Since the population densities and human mobility indices of rural regions are very different from those of cities, it is imperative to develop specific mitigation schemes to impede the spread of epidemics right from its likely source, the rural location. Furthermore, recent studies show that rural residents have a lower likelihood to obtain certain preventive health services than urban residents \cite{CCK}. These factors necessitate research on predictive and optimally preventive strategies in rural regions.

This paper takes a unique look at rural regions, and presents mitigation strategies tailored for rural Clay county in Kansas, USA. We propose mitigation strategies that are based on a contact network model developed using data collected through a survey campaign conducted in rural Clay and Kearny counties in Kansas, USA. By characterizing the contact structure of rural regions, we are able to investigate the influences of this structure and various mitigation strategies on the speed, shape, and size of the outbreak. Our analysis shows that, although global targeted strategies are the most efficient in mitigating epidemics with a limited amount of resources, they can also be unfeasible due to partial knowledge of the population and conflicts with individual rights. Random vaccine distribution in selected popular location within a rural community offers the opportunity to indirectly reach the individuals who play a significant role in the epidemic propagation. We demonstrate with simulations that this location-based strategy can be 55\% as effective as the best global target strategy.

\section{Methods} 

Our simulation results on epidemic spreading in rural communities are based on data collected through distributed surveys. This data is used to construct a contact network and to analyze the epidemic. Several random and targeted mitigation strategies are investigated through an SEIR model with parameters estimated in an analysis of the recent H1N1 influenza \cite {V}. 

\subsection{Survey Data}

In Spring 2009, we surveyed residents of two rural Kansas counties through a visit to a county seat and mailed surveys, under our direct personal supervision. We obtained ethics approval in January 2009 for research protocols, survey forms, and informed consent procedures used, by the Institutional Review Board (IRB) of Kansas State University, Human Subjects Committee, University research Compliance Office, 203 Fairchild Hall, Manhattan, Kansas. All potential participants were provided informed consent in a cover letter attached to their surveys; signatures were not required from the participants as a way of protecting their privacy.  

The mailed surveys were well accepted with response rates of 64.8\% and 41\%, respectively for Clay County and Kearny County. The survey consisted of 30 short questions, a question concerning visits to local businesses and locations, a question concerning visits to cities within the surrounding region, and a set of contact questions. The spread of an epidemic in rural areas may be influenced by both the vulnerability of the population and the extent of their contacts with each other. Vulnerability includes their susceptibility to infection due to both poor health and a lack of preventive measures, such as vaccination. Once an epidemic has begun, the willingness of the population to comply with precautionary health measures can influence the rate and extent to which the epidemic spreads. In the survey, all these factors were assessed. 

Survey results yielded four measures of risk factors important to the spread of epidemics: health risk, contact risk, prevention risk, and compliance risk. To what extent did these risks overlap? Each possible combination of risks was evaluated and summarized in {Table \ref{table1}}. It is interesting to note that people with the most contacts tended to have the least preparedness for an epidemic, and people who were willing to visit others even during an epidemic were among those most at-risk because of their health status. Additionally, those who tended to visit friends and family members more often during normal times were also likely to retain this behavior even under epidemic conditions. This property of the rural communities is interesting, since it provides a given level of stability within the contact network and increases the accuracy of our epidemic analysis.

Survey results were also used to construct the weighted contact network. To this purpose, we used the survey responses about frequently visited locations and the levels of contacts. The respondents were asked to identify within a set of locations, those which they visit on a typical day. The responses were captured as a binary vector $L_{i}$ for each respondent $i$ with each element corresponding to a location. The contact questions asked the respondents to estimate the number of individuals with whom the respondent made contact for three different levels of contact. Contact levels were classified into \textit{Proximity} contact (coming within 5 feet of another person, even if in passing), \textit {Direct-Low} contact (directly touching another person for a short period of time in what most people would consider a low risk situation of being infected), and \textit {Direct-High} contact (directly touching another person for an extended period of time or in what most people would consider a relatively high risk situation of being infected). The responses of the contact questions are quantified as values $n_{x,i}$ that represent the number of individuals contacted by respondent $i$ in a typical day according to each contact level $x$.

\subsection{Contact Network Construction}

With these responses, the rural community is represented as a weighted contact network where each of the survey respondents is represented as a node within the network that is connected together with links representing the contact between respondents. Each link has a weight that represents the normalized measure of contact between the connected pair of respondents or nodes. Each link's weight $w_{i,j}$ is taken as the average of three sub-weights that correspond to the interactions between node $i$ and node $j$ estimated for each contact level $x$. Values of weights $w_{i,j}$ range within the interval [0, 1]. We capture the location responses within the parameter $\mu_{i,j} = (1+l_{i,j})/(1+d)$, where $d$ is the total number of locations and $l_{i,j}$ is the dot product of the respective location vectors $L_{i}$ and $L_{j}$ for nodes $i$ and $j$. 
For a give type of contact $x$, the related sub-weight function depends on the node degrees and the parameter $\mu_{i,j}$. When either node degree  $n_{x,i}$ or $n_{x,j}$ is zero, the sub-weight should be equal to zero. The sub-weight should also increase monotonically with both $n_{x,i}$ and $n_{x,j}$, approaching unity when both are large. When a pair of nodes visit all the locations,  $\mu_{i,j}$ is equal to unity and the sub-weight should be maximum. On the other hand, $\mu_{i,j}$ has a small positive minimum, to allow for interactions outside the locations included in the survey. For given $n_{x,i}$ and $n_{x,j}$, the sub-weight should be minimum when $l_{i,j}$ is equal to zero, and should increase monotonically with increasing $l_{i,j}$, and consequently $\mu_{i,j}$.

For each contact level $x$, we compute the sub-weight $w_{x,i,j}$ between node $i$ and node $j$ according to a simple function which follows the desired behavior:

\begin{equation}
\label{equation1}
w_{x,i,j} = (1 -(1 - \mu_{i,j} \pi_{x})^{n_{x,i}})(1 - (1 - \mu_{i,j} \pi_{x})^{n_{x,j}})     
\end{equation}

We selected the values of $\pi_{x}$ such that they include the relative importance of each of the three contact categories from the survey responses, constraining $\pi_{Proximity}$ to be less than $\pi_{Direct-Low}$ and $\pi_{Direct-Low}$ to be less than $\pi_{Direct-High}$. The values of $\pi_{x}$ have been estimated by matching the epidemic curve of the H1N1 outbreak in La Gloria, Mexico, with the average epidemic curve obtained with the weighted network simulations for Clay Center. Based on the minimum squared error, we found that the best contact levels for Proximity, Direct-High and Direct-Low contact are $\pi_{Proximity} = 0.0025$, $\pi_{Direct-Low} = 0.015$, and $\pi_{Direct-High} = 1.0$. {Figure \ref{nomitigation}} reports the number of new infected individuals in La Gloria \cite{F} with the corresponding simulated new infected individuals in Clay Center, Kansas given the best estimated values for the three contact levels. This parameter estimation was done through an SEIR epidemic model, which we describe in the following sections.

With the estimated model parameters, we have created the weighted contact network shown in {Figure \ref{A1new}-\ref{B3new}}. In {Figure \ref{A1new}}, only links with strength greater or equal to 0.2 are depicted. In the successive subfigures more links are added by reducing the threshold, reaching the complete network in {Figure \ref{A6new}}, where all links are depicted. {Figure \ref{B3new}} is the union of two networks: the network obtained for threshold 0.1 and the network of the \textit{best friends} where for each node only the link with highest strength is depicted. {Figure \ref{B3new}} has been used to create {Figure \ref{newclaycenternew}}, where not only nodes representing people, but also nodes representing popular locations in Clay Center are shown. We performed a sensitivity analysis on each $\pi_{x}$, varying their values up to 15\%. These variations, shown in {Table \ref{table2}}, produce a maximum of 3.4\% variation in total infection cases, with most changes resulting in a variation of less than 1\% in total cases. 

To compare our selection of the structure of the weight $w_{x,i,j}$ shown in Eq. \ref{equation1}, we have constructed another weighted contact network based only on data about common visited locations. In this case, the weights $w^1_{i,j}$ are computed as $w^1_{i,j}=(l_{i,j})/(d)$. The network constructed in this way has 38\% of nodes isolated, i.e., with node degree equal to zero, and when we simulated the SEIR model only 63\% of the infected nodes coincided with the infected nodes obtained using the same model on our contact network. Consequently, the use of only location data produces not negligeable differences in the results and is not considered in the following analysis.

\subsection{Network Metrics} 

To describe in details the characteristics of the weighted contact network, we select some graph-theoretical metrics that reflect the local and global properties of the graph \cite{BBV}. In {Table \ref{table3}}, some relevant metrics for the contact networks are listed. The contact network is composed of 138 nodes ($N$) and 9222 links.  It is important to note that that the network is not far from a fully connected network, which would have 9453 links. However, each link can have a very different importance due to the structure of the link weights. For this reason, we select the node strength as one metric to characterize a node. The strength $s_{i}$ of node $i$ is defined as the sum of the weights $w_{i,j}$ of all links between node $i$ and its neighbors $s_{i}= \sum_{j \in neighbors(i)} w_{i,j}$. The node strength is analogous to the node degree in the binary network, which measures the number of contacts or neighbors of a node.
The second metric we compute is the average shortest path. To compute shortest path properly, we define the distance $d_{i,j}$ between any neighbor nodes $i$ and $j$ as $d_{i,j} = 1-w_{i,j}$. The distance defined in this way is always non-negative and reveals a short distance separating node $i$ and node $j$ when their link weight $w_{i,j}$ is high. The third metric, the network diameter is defined as the longest of the shortest paths.
As a centrality measure, we compute the betweenness $b_{i}$ of a node $i$. It is defined as the measure of the number of shortest paths between any pair of nodes passing through node $i$.
\begin{equation}
b_{i} = \sum_{ h,j} \frac{\sigma_{h,j,i}}{\sigma_{h,j}}
\end{equation}
where $\sigma_{h,j}$ is the total number of shortest paths from node $h$ to $j$ and $\sigma_{h,j,i}$ is the number of those shortest paths that pass through the node $i$. A node that appears in many shortest paths has high betweenness. Each $b_{i}$ is normalized by the maximum number of shortest paths that can pass through a node $(N-1)*(N-2)/2$.
Another measure of node centrality is the clustering coefficient of a node $i$ $c_{i}$, which measures the level of connection among the neighbors of node $i$.
\begin{equation}
c_{i} = \frac{1}{k_{i} (k{i}-1)} \sum_{j,k} [\hat{w}_{i,j} \hat{w}_{i,k} \hat{w}_{j,k}]^{\frac{1}{3}}
\end{equation}
where $\hat{w}_{i,j} = w_{i,j}/max(w_{i,j})$ and $k_{i}$ is the degree of node $i$ in the binary version of the weighted contact network \cite{SKOKK}. By averaging over all individual clustering coefficients, we obtain the average clustering coefficient of the contact network. 
The node coreness is the maximum value $k$ such that the node still exists in the network, before being removed in the $k+1$ core.  The $k$-core of a graph is a maximal subgraph in which each vertex has at least strength $k$. The coreness measures the deepness of a node in the core of the network where a higher value indicates that the node is deeper in the core. The discrete values for the strength classes are obtained by a fine quantization of the node strength (step size on the order of $10^{-6}$). From the spectral domain, the maximum eigenvalue is the largest eigenvalue of the weighted adjacency matrix $W$ representing the network. The elements of $W$ are the weights $w_{i,j}$, and the matrix in this case is symmetric and has zeros in the main diagonal. A large maximum eigenvalue corresponds to a small epidemic threshold in the Susceptible-Infected-Susceptible model \cite{SSGE}. 

Networks often display some level of grouping of nodes in an organized fashion that allow them to be divided into different clusters or communities. One popular method of detecting communities is to maximize a parameter known as modularity \cite{NG}. Modularity is a measure of the difference between the edges within each community and the expected number of edges in the same community, summed over all communities within the graph. Using the weighted version of the algorithm described in \cite{N}, we found two communities within our contact network. With a modularity value of 0.1087, 61\% of the population fell into community 1 and the remaining 39\% of the population composed community 2.

\subsection{SEIR Model on the Contact Network}

We expand the weighted compartmental model \cite{SSGE} to represent the different disease states of individuals: Susceptible, Exposed, Infectious, and Recovered. The states and the transitions between states are unique to each disease and its characteristics, requiring customization to each disease. In this model, we selected $\beta = 0.4$, where $\beta$ is the rate of infection across a link between a susceptible individual and infected individual, $\epsilon \approx 0.909$, where $\epsilon$ is the transition rate parameter between the Exposed and Infected compartments, and $\delta = 0.4$, where $\delta$ is the transition rate parameter between the Infected and Recovered compartments \cite{KM}.
The network topology plays an important role in the spreading process in the transition from $S$ to $E$ when an $I$ individual contacts an $S$ individual and successfully infects him/her. The probability that node $i$ is not infected at time $t$ depends on the probabilities that a neighbor node $j$ is previously infected ($p_{j,t-1}$), is in contact with node $i$ ($w_{i,j}$), and successfully infects node $i$ ($\beta$) \cite{SSGE}.
\begin{equation}
\zeta_{i,t}= \prod_{j \in contacts} (1- w_{i,j} \beta P_{j,t-1})
\end{equation}
The probability that a node is infected (transition from $S$ to $E$) at time $t$ is then $1 - \zeta_{i,t}$. The remaining transitions are topology independent and only depend on the rate parameters, $\epsilon$ and $\delta$, of the disease model. When an individual has contracted the disease and transitioned into the exposed or latent compartment, the individual transitions to the next state ($I$) with rate $\epsilon$. Once Infected, a node attempts to infect it's susceptible neighbors until it transitions to the recovered state. Each Infected node recovers with recovery rate $\delta$. Once a node is recovered ($R$), it remains recovered for the remainder of the simulation. The recovered compartment serves as an accumulator of all the cases, thus the number of recovered individuals $\vert R\vert$ at the end of a simulation is a decent approximation of the total number of cases caused by the outbreak. The blue curve in {Figure \ref{nomitigation}} has been computed using the above model.

\subsection{Epidemic Simulations}

The analysis of the epidemic evolution and the evaluation of multiple mitigation strategies is performed using an SEIR model on the contact network. We propose different immunization strategies that can be implemented as vaccinations or antiviral treatments. The immunization strategies are classified in three categories based on individuals, locations, and communities. In each individual immunization strategy, nodes are chosen either deliberately, based on a node metric or randomly.

The random selection of nodes as recipients of an immunization represents an unbiased distribution of resources and is the simplest method for distribution. The node metrics selected for the targeting strategies include node strength, node coreness, and node betweenness. Node strength, as a measure of how well an individual is connected with the rural population, is an intuitive measure of how likely a node is to be infected by other nodes as well as how likely the node is to pass the infection on to others. Therefore to mitigate the infection while using node strength to select nodes, we target the nodes with the highest strength. The node coreness is a measure of how \textit{deep} a node is in the core of a network. This depth is a measure of the maximum strength of the nodes iteratively removed from the network periphery before the node is removed. From a topological perspective, the core of the network facilitates connectivity and is vital for it. Therefore a targeted removal or immunization of the core nodes serves to hinder and disrupt the connectivity that allow the spread of the infection. The betweenness of a node measures how many shortest paths between all pairs of nodes choose to route through the node. Thus targeting nodes with highest betweenness serves to disrupt the shortest paths that the virus can take, forcing it to longer routes. We applied these different targeting strategies globally on the entire network and then within the communities and selected locations. The immunization of a node is implemented by forcing the immunized nodes to remain Susceptible throughout the epidemic.

In {Table \ref{table4}} the reduction in the number of cases by percentage with respect to the unmitigated epidemic is shown, for different criteria for the selection of the 10\% of immunized people among the global population. The most effective strategy is the one where the 10\% of nodes with highest strength are selected, in line with previous results, followed by the one based on the selection of 10\% of the nodes with the deepest coreness. However, these types of strategies have an inherent problem: how can we practically detect those special nodes? Fortunately, the data collected on the location popularity, can help to solve this problem. The survey respondents associated themselves with various locations in the county by indicating which ones they typically visit. We used two criteria to select the locations for targeting and random strategies. To select locations for the random strategies, we chose the locations having the highest average value of the desired metric and being associated with at least 10\% of the population. For the targeting strategies, we chose the locations visited by more than 10\% of the population, and we immunized 10\% of the population by selecting nodes with the highest combined sum for the desired metric within those locations. In {Table \ref{table5}}, the reduction in the number of cases by percentage is shown, when the immunization of the selected people is performed among the group visiting a particular location.

\section{Results and Discussion}

Obtained results span the two investigated areas, namely risk assessment and mitigation strategy evaluations. Concerning risk assessment, very few rural respondents (2\%) did not have a high level of risk in at least one of four areas assessed: health risk, contact risk, prevention risk, and compliance risk.  Over 75\% of households did not have complete uptake of flu vaccine, nearly half of respondents had at least one major health risk, and nearly two-fifths of respondents said they would not comply with directives to stay at home during an epidemic. Risk levels were positively associated, suggesting that risks were compounded with each other, a situation posing greater problems for any attempt to predict or reduce the spread of epidemics in rural areas. Married respondents were much less likely to report selected health risks by a substantial margin (38\% vs. 69\%).  Other demographics factors had relatively small associations with health, compliance, prevention, or contact risks, although some nonlinear associations between income and the risk factors were noted, with middle-income respondents having the lowest risk levels compared to lower or higher-income respondents.

Concerning mitigation strategies evaluation, {Table \ref{table4}} shows that the random immunization of 10\% of the population (first strategy) reduces the epidemic size by 11.40\%, with no substantial gain. However, if 10\% of the nodes with highest node strength are immunized (second strategy), the epidemic size is reduced by 34.57\%, more than three times the size of the random immunization campaign. In the interesting case where the 10\% of the immunized nodes are randomly selected within the group of people frequently visiting a specific popular location (third strategy), an intermediate benefit, of about 19\% epidemic size reduction, is obtained. The identification of specific locations visited by highest strength nodes has the clear benefit of improving the efficiency of a random immunization campaign, when this campaign is conducted in specific locations.  {Figure \ref{results}} shows the curves of new infected nodes with time under free evolution and for the discussed three mitigation strategies.

Our simulations suggest that information and immunization activities for rural communities should be carried out in specific locations, called key locations, which not only most people but also the most key people (highest strength nodes) often visit. Detecting key locations requires some amount of data collection and analysis. However, detecting key locations is much easier than identifying highest strength nodes. In other words, the probability of immunizing a highest strength node given a node random selection in a key location is much higher that the probability of immunizing a highest strength node given a node random selection in the entire population.

In the presence of limited anti-viral and vaccination resources, government health agencies should seek to use the most effective methods of distribution for mitigation of the threat. Here we have investigated the distribution of immunizations to 10 percent of the population through various targeting strategies. This work is of particular interest to rural regions as they are more likely to face resource shortages due to smaller budgets than urban areas. Due to lower population densities, rural regions are more likely to have a small set of local businesses and locations that are therefore easier to classify and target for distribution. This work has shown the benefit of being able to select proper distribution locations, a strategy that can be implemented without having full knowledge of every individual within the rural population.



\begin{table*}
\centering
\caption{Interrelationships of risk measures}
\begin{tabular*}{\hsize}{@{\extracolsep{\fill}}lll}\hline
Health-Contact & Health-Prevention & Health-Compliance\cr \hline
\hline
As contact risk rose from low to & Of those respondents from families in & The percentage of respondents  \cr medium to high, the percentage of & which all members had been & with one or more at-risk health  \cr respondents with one or more & vaccinated, only 34.1\% had one or & conditions tended to rise as a  \cr 
at-risk health conditions rose & more at-risk health conditions, & function of their unwillingness \cr 
linearly from 37.7\% to 45.0\% to & compared to 49.6\% of those from &  to comply with a directive  \cr 
56.3\% (p $<$ 0.18 by chi-square & families in which at least some of the & to stay at home during an  \cr 
test; $r$= 0.15, $p$ $<$ 0.07) & members had not been vaccinated &epidemic : no visits (46.1\%),  \cr
&(p $<$ 0.09).& one or two (38.8\%), and three\cr
& & or more(75.0\%)(p $<$ 0.08).\cr 
\hline
Those who would be most & Those who were most susceptible to & Those who were willing to visit\cr 
vulnerable or susceptible to an & an epidemic were least likely to be &  others even during an epidemic \cr 
epidemic were actually most & prepared for it in terms of anticipatory & were among those most at-risk \cr 
likely to be engaging in multiple & vaccination. & because of their health status. \cr 
contacts with friends, family, & &  \cr 
and guests. & & \cr
\hline
Contact-Prevention & Contact-Compliance & Prevention-Compliance \cr\hline
\hline
As contact levels rose from low to & As contact levels rose from low to & Those from families that were \cr
medium to high, the percentage of & medium to high, the percentage of & fully vaccinated declined  \cr
households with full vaccinations & respondents who would not comply & from 29.1\% to 22.0\% to 0.0\% as \cr 
fell from 32.1\% to 19.7\% & with health directives to remain at & respondents shifted away from  \cr 
and 20.8\%, respectively ($p$ $<$ 0.25 & home rose from 29.4\% to 40.0\% to & no visits, one or two visits,\cr 
by chi-square test; $r$ = 0.11, & 44.7\%, respectively ($p$ $<$ 0.28 by &  or three or more visits during an \cr 
$p$ $<$ 0.18). & chi-square test; $r$ = 0.13, $p$ $<$ 0.12). &  epidemic ($p$ $<$ 0.08 by chi-square \cr
& &test; $r$ = 0.17, $p$ $<$ 0.04). \cr
\hline
Those with the most contacts & Those who tended to visit friends and & Those making the most visits \cr 
tended to have the least & family members more often during & were the least likely to be  \cr 
preparedness for an epidemic. & normal times were also likely to & protected by vaccination. \cr
& retain this behavior even under &  \cr
& epidemic conditions. &  \cr 
\hline
\end{tabular*}
\label{table1}
\end{table*}

\begin{table}
\centering
\caption{Sensitivity analysis showing percentage differences from original number of total cases}
\begin{tabular*}{\hsize}{@{\extracolsep{\fill}}ccccccc}
\hline
 Percentage Difference& -15\% & -10\% & -5\%	& +5\% & +10\% & +15\% \cr\hline
\hline
$\pi_{Proximity}$ &	-0.29334 & -0.22053	& -0.10083 & -0.00539 &	0.15957 &	0.22018 \cr
$\pi_{Direct-Low}$ & -0.70028	& -0.49687 & -0.26954 &	0.20365 &	0.50449 &	0.77051 \cr
$\pi_{Direct-High}$ &	-3.44748 & -2.02815 &	-0.95024 & X & X & X \cr
\hline
\end{tabular*}
\label{table2}
\end{table}

\begin{table}
\centering
\caption{Network Metrics for Contact Network}
\begin{tabular*}{\hsize}{@{\extracolsep{\fill}}lclc}
\hline
Network Metric&Value&Network Metric&Value\cr \hline\hline
Links & 9222 & Aver. Link Weight & 0.006454 \cr
Diameter in no. of Hops & 2 & Aver. Node Coreness & 0.49579034 \cr
Aver. Clustering Coefficient & 0.0037 & Aver. Node Betweenness & 0.000179682 \cr
Aver. Node Strength & 0.8626 & Aver. Shortest Path By Distance & 1.01699 \cr
Diameter by Distance & 1.99996 & 	Max Eigenvalue of Weighted Matrix & 480.5959 \cr

\hline
\end{tabular*}
\label{table3}
\end{table}



\begin{table}
\centering
\caption{Global Mitigation Strategies}
\begin{tabular*}{\hsize}{@{\extracolsep{\fill}}lccc}
\hline
Mitigation Strategy 10\% Immunization &	\% Reduction of Total Cases & Cases Prevented per Vaccine \cr \hline
\hline
Random &	11.40 &	0.69 \cr
Highest Strength Nodes	& 34.57 &	2.11 \cr
Highest Coreness Nodes	& 25.18 &	1.53 \cr
Highest Betweenness Nodes &	16.27 & 0.99 \cr
\hline
\end{tabular*}
\label{table4}
\end{table}

\begin{table}
\centering
\caption{Location-based Mitigation Strategies}
\begin{tabular*}{\hsize}{@{\extracolsep{\fill}}lccc}
\hline
Mitigation Strategy 10\% Immunization	& \% Reduction of Total Cases &	Cases Prevented per Vaccine \cr \hline
\hline
Random in Most Popular Location & 14.25 &	0.87 \cr
Random in Highest Strength Location &	18.97 &	1.16 \cr
Highest Strength in Most Popular Location &	34.57 &	2.11 \cr
Highest Coreness in Most Popular Location &	25.17 &	1.53 \cr
Highest Coreness in Highest Coreness Location	& 24.31 &	1.48 \cr
\hline
\end{tabular*}
\label{table5}
\end{table}

\begin{figure*}
\centerline{\includegraphics[width=.5\textwidth]{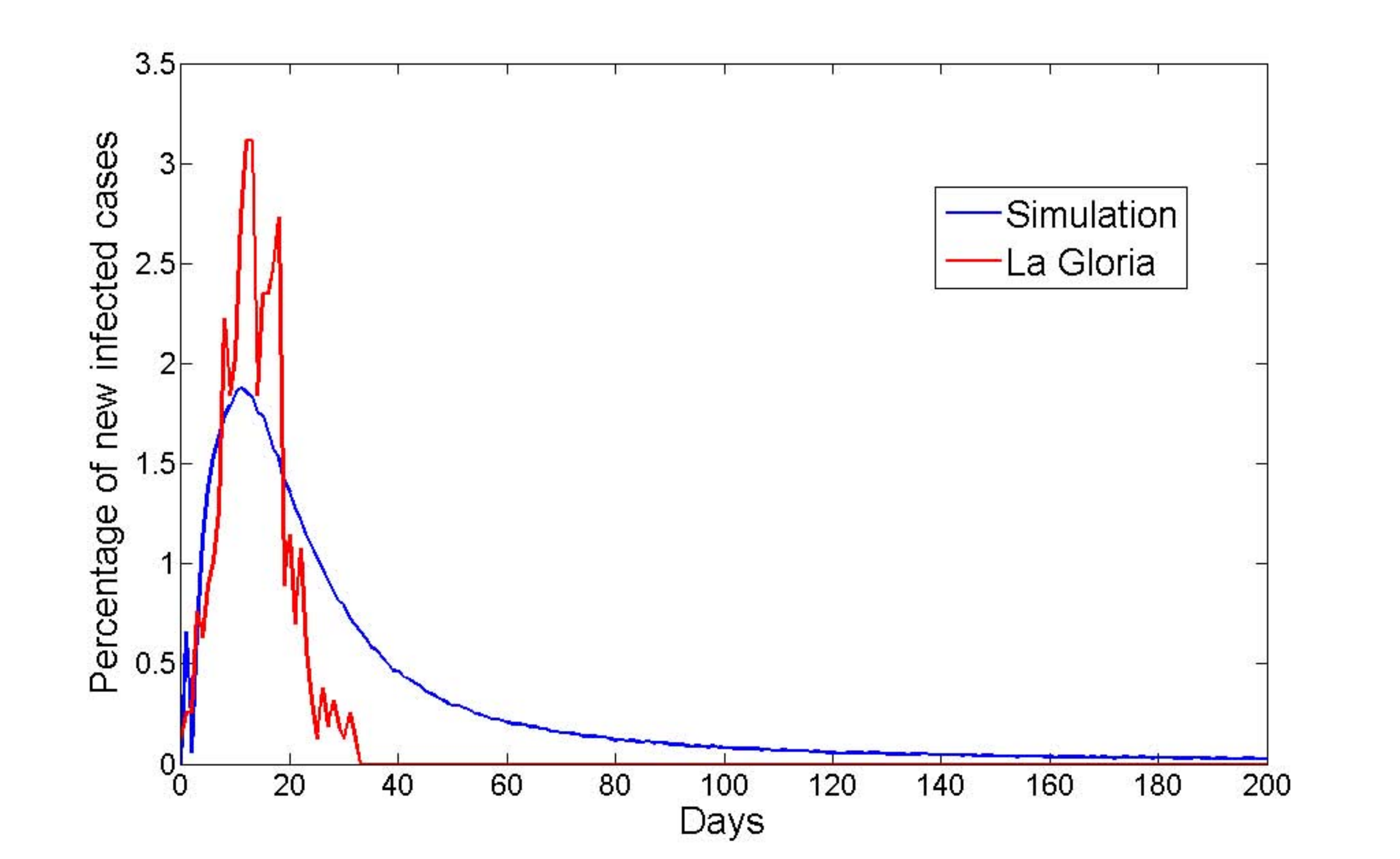}}
\caption{Number of new infected individuals in La Gloria, Mexico by day, \cite{F} with the corresponding new infected individuals in Clay Center, Kansas, given the best estimated values for the three contact levels, averaged over 10,000 simulations.}\label{nomitigation}
\end{figure*}

\begin{figure*}[ht]
  \begin{center}
    \subfigure[link threshold 0.2]{\label{A1new}\includegraphics[scale=0.2]{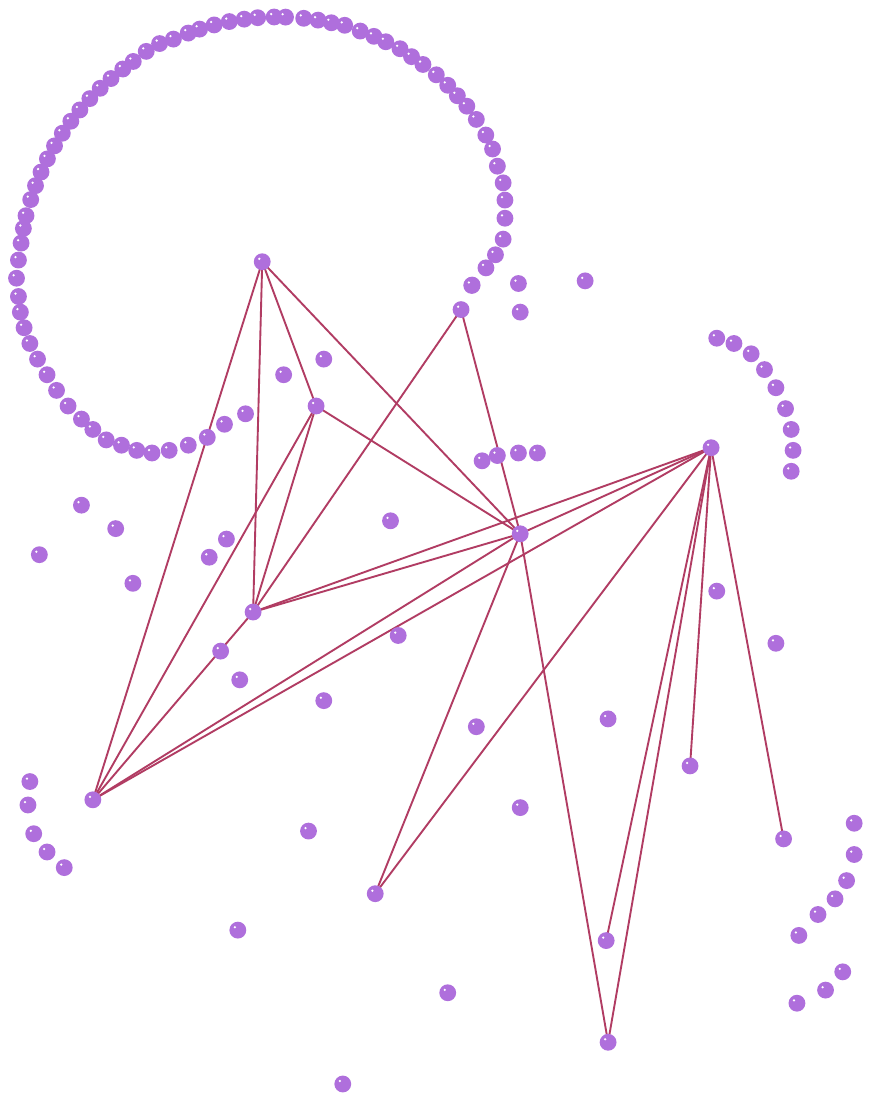}}
    \subfigure[link threshold 0.1]{\label{A2new}\includegraphics[scale=0.2]{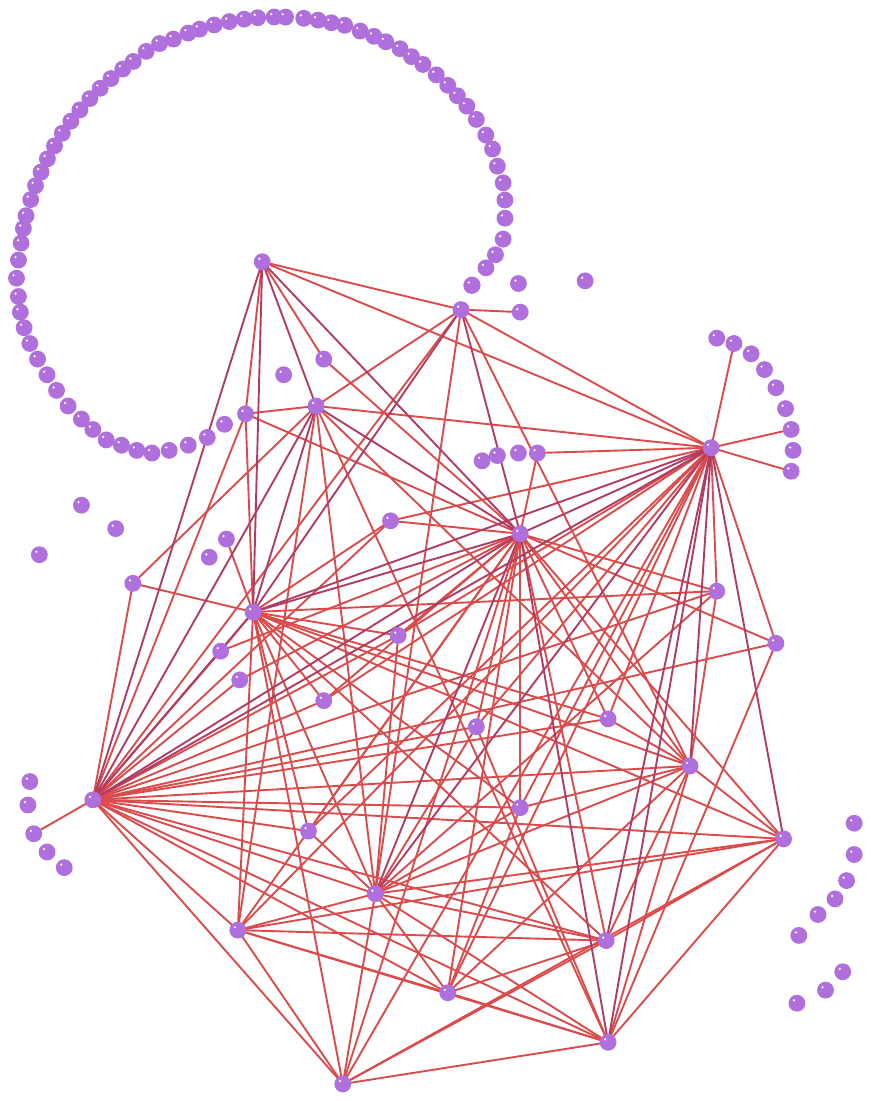}}
    \subfigure[link threshold 0.05]{\label{A3new}\includegraphics[scale=0.2]{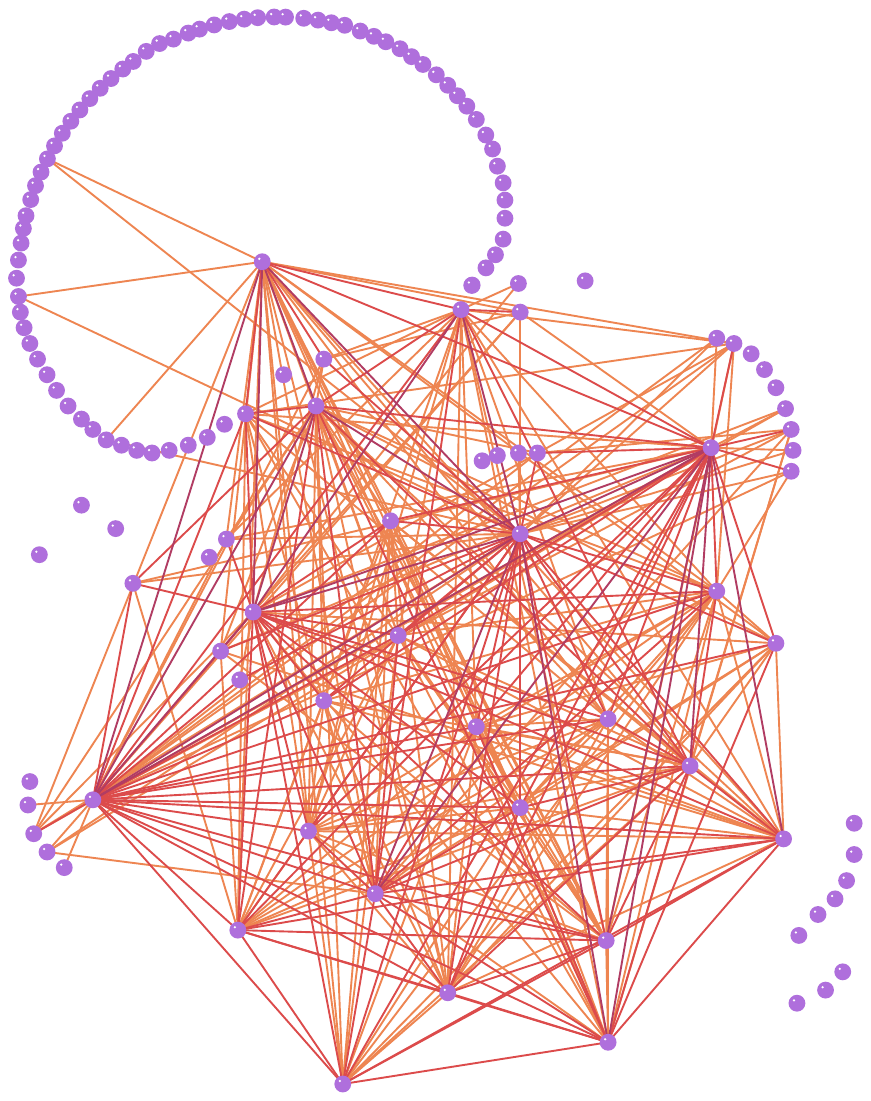}}
    \subfigure[link threshold 0.0125]{\label{A4new}\includegraphics[scale=0.2]{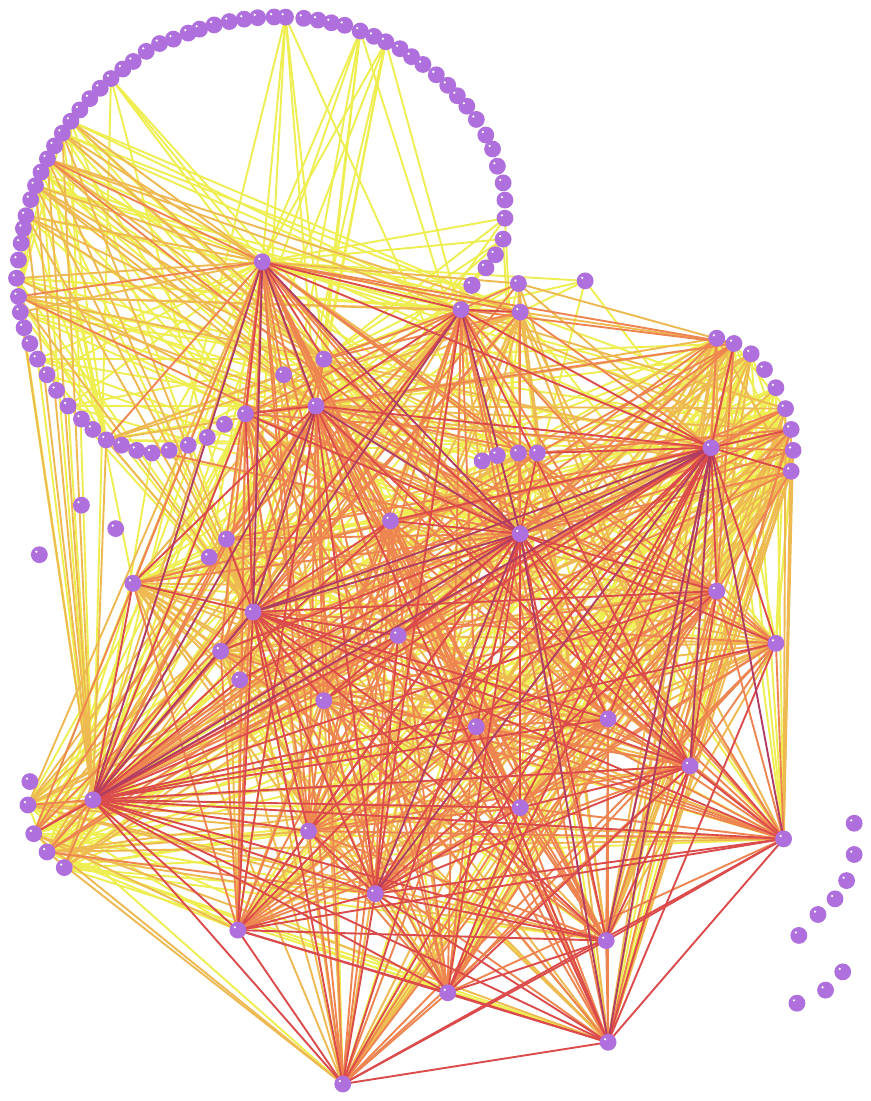}}
    \subfigure[link threshold 0.003125]{\label{A5new}\includegraphics[scale=0.2]{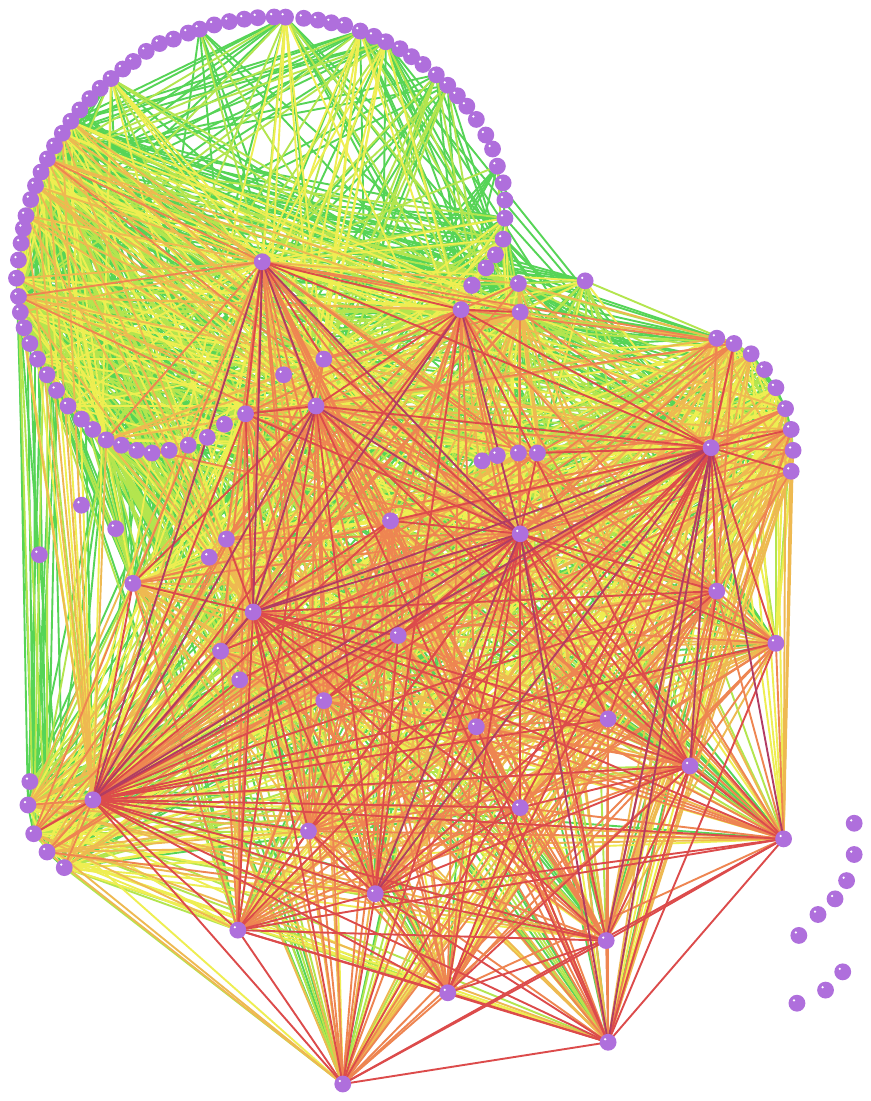}}
    \subfigure[complete network]{\label{A6new}\includegraphics[scale=0.2]{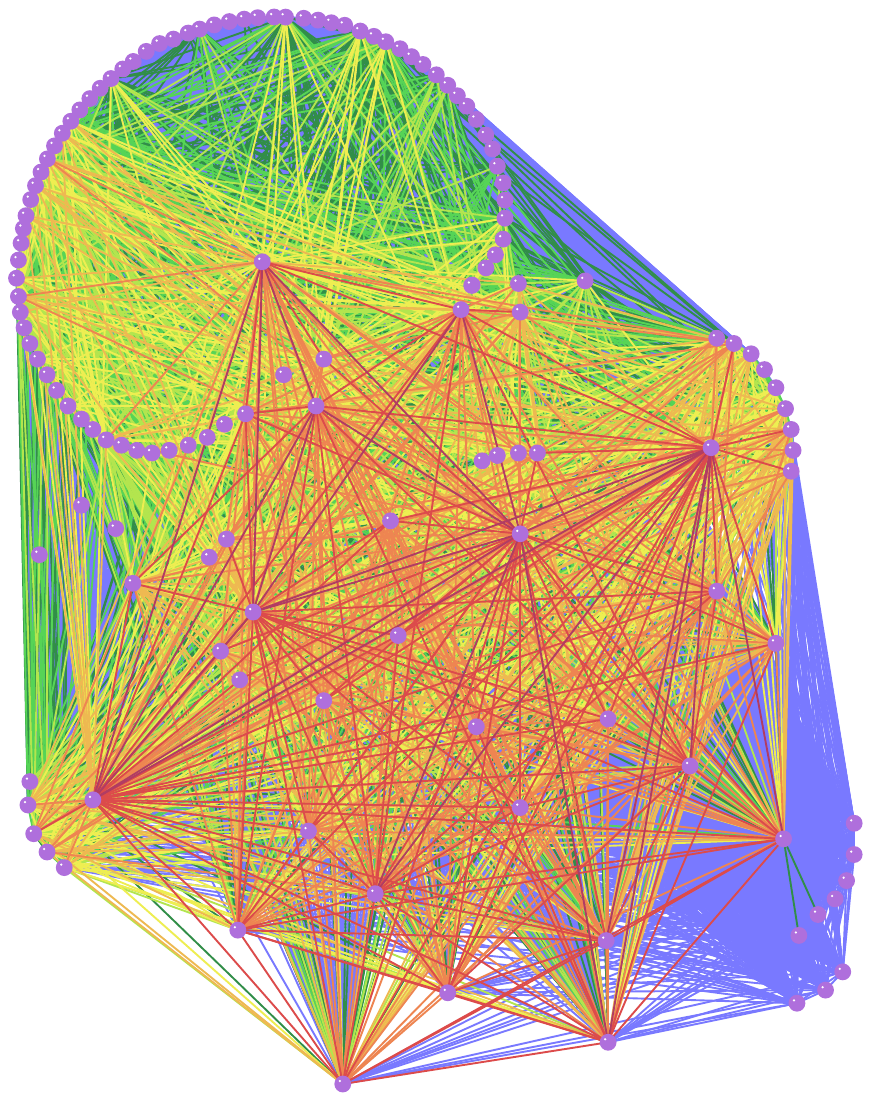}}
    \subfigure[link threshold 0.1]{\label{B1new}\includegraphics[scale=0.2]{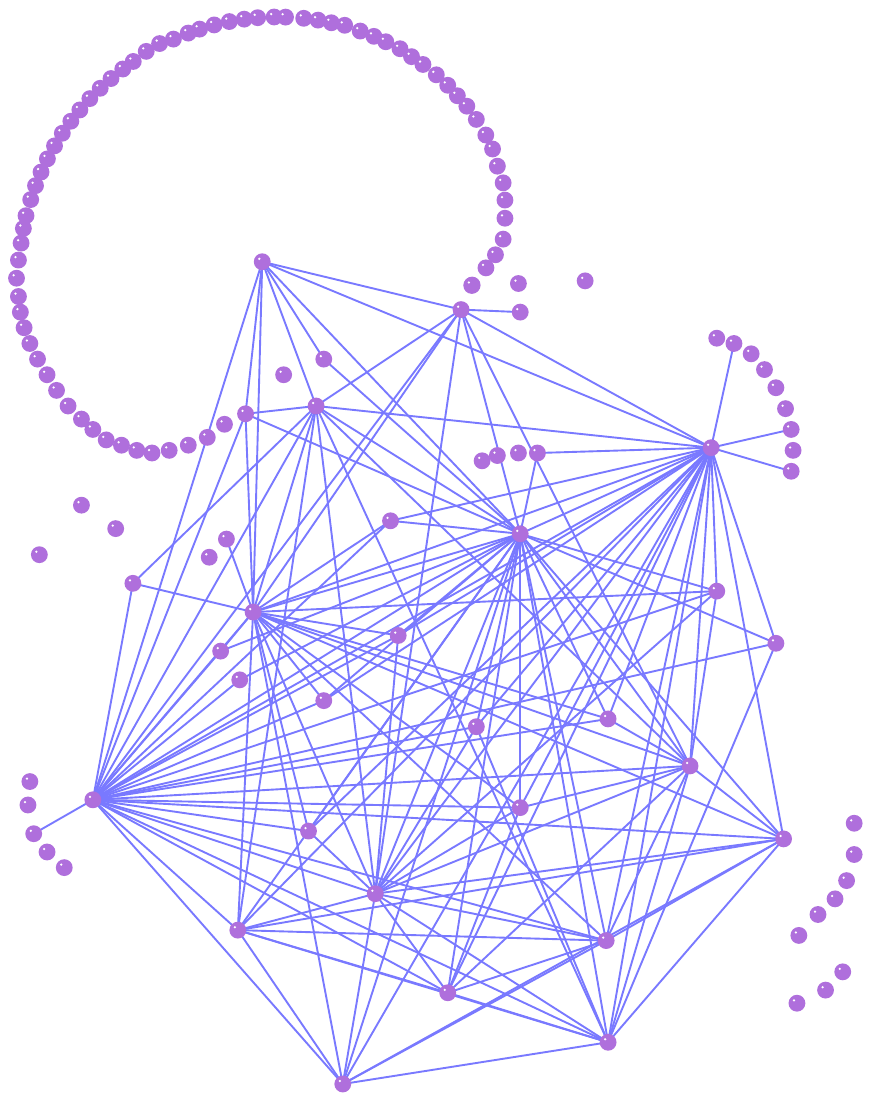}}
    \subfigure[\textit{best friend} network]{\label{B2new}\includegraphics[scale=0.2]{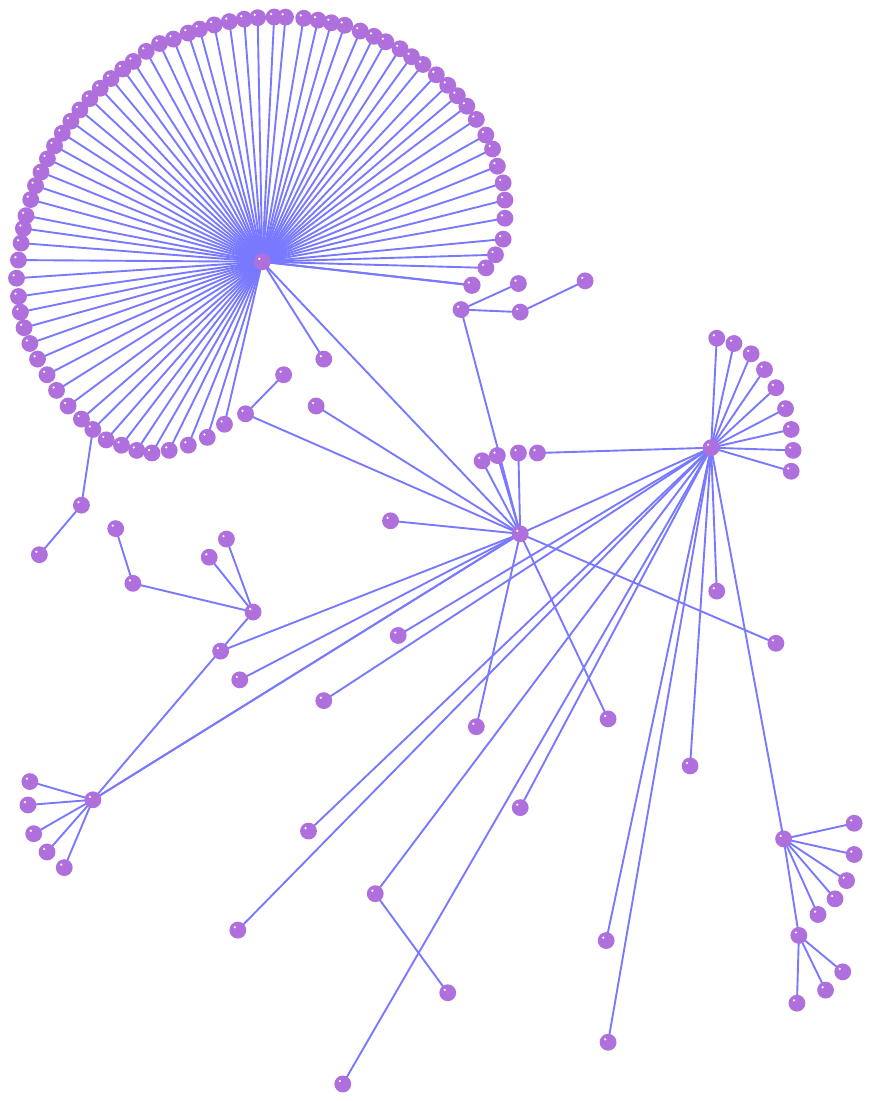}}
    \subfigure[union of \ref{B1new} and \ref{B2new}]{\label{B3new}\includegraphics[scale=0.2]{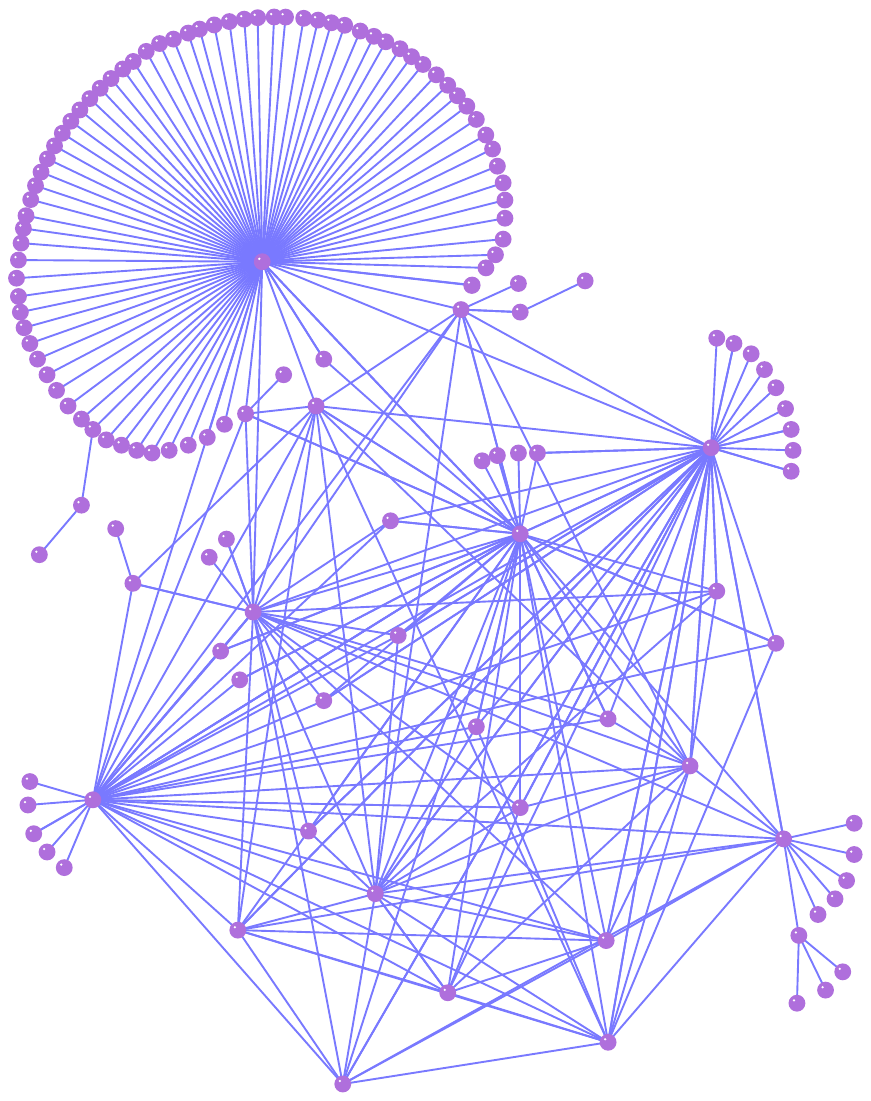}}
  \end{center}
  \caption{The rural contact network is composed by nodes representing individuals and weighted edges representing contacts, displaying all edges with weights greater than the following thresholds: a) 0.20, b) 0.10, c) 0.05, d) 0.0125, e) 0.003125, and f) 0. In the last row, g) shows the edges with weights greater than 0.10, h) shows the highest weighted edge for each node, and i) shows the union of the previous two networks. }
  \label{fig:edge}
\end{figure*}

\begin{figure*}[ht]
\begin{center}
\centerline{\includegraphics[height=8.0cm,width=12.0cm]{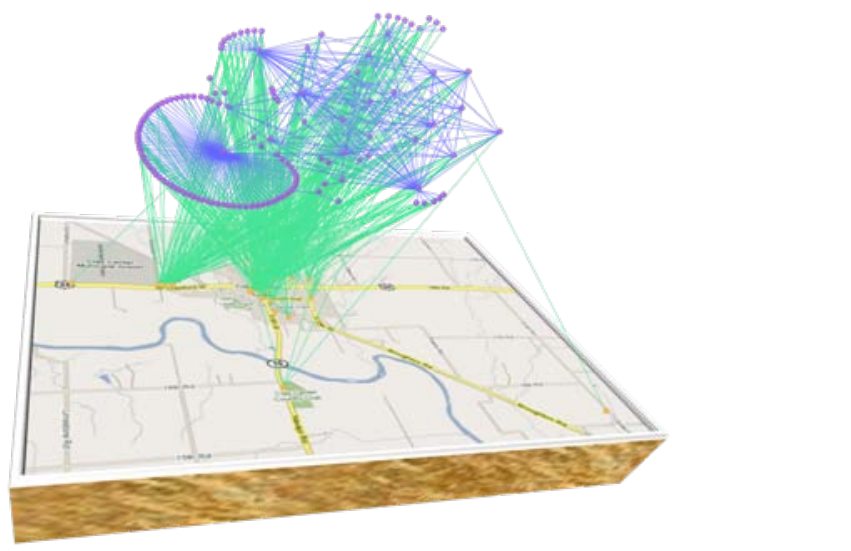}}
\caption{The network of people and popular locations in Clay Center, Kansas, where the nodes (survey respondents) in the cloud network are connected via green edges to the locations in Clay Center according to the survey responses. The map is courtesy of Google.}
\label{newclaycenternew}
\end{center}
\end{figure*}

\begin{figure*}
\centerline{\includegraphics[width=.5\textwidth]{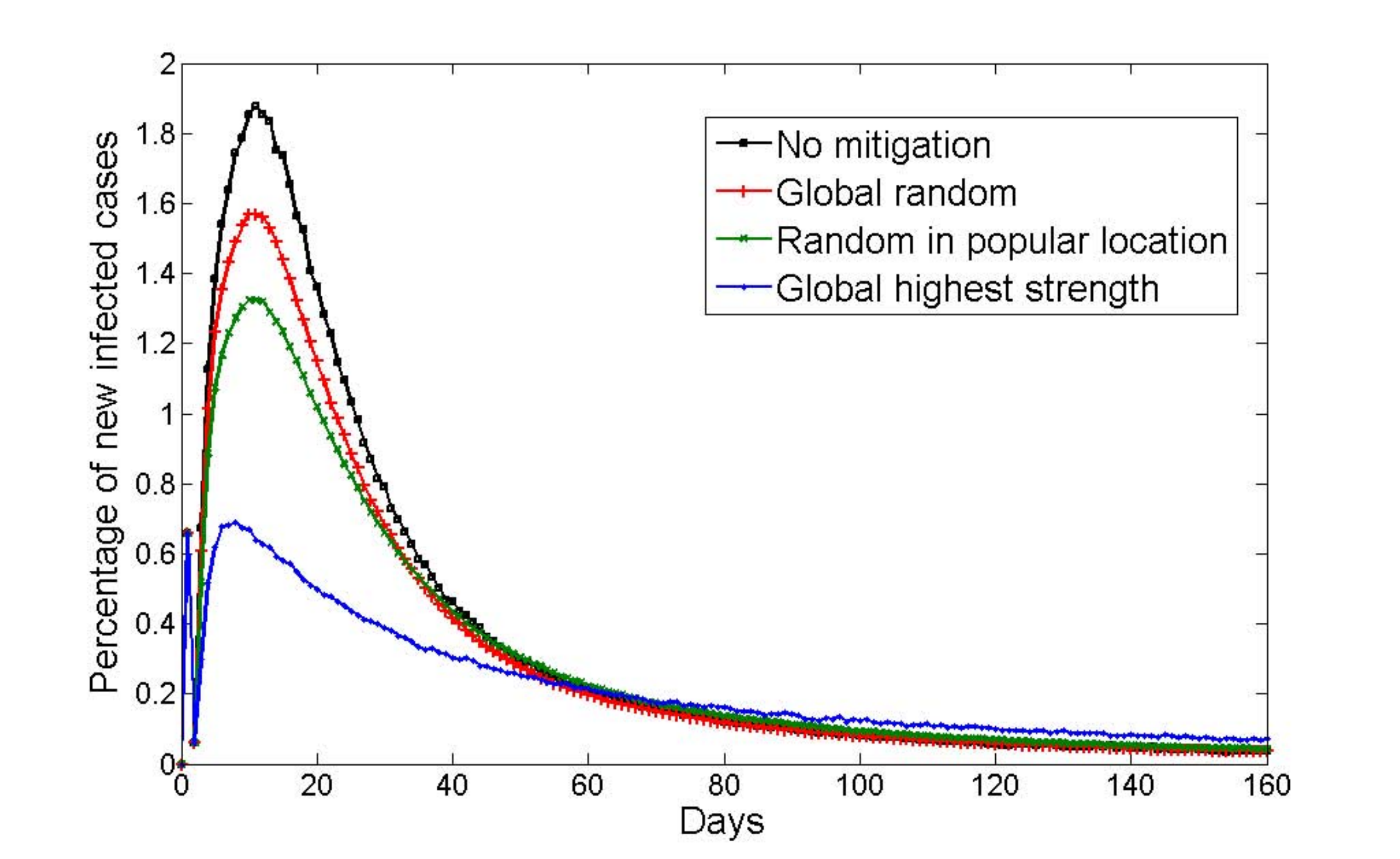}}
\caption{Newly infected nodes by day as a percentage of the population without and with mitigations strategies, including random vaccination throughout the population, random vaccination among nodes associated with a selected location, and targeted vaccination of a set of nodes having the highest node strength within the rural contact network.}\label{results}
\end{figure*}

\end{bmcformat}
\end{document}